# Scalable Quantum Computing With "Enhancement" Quantum Dots

Y. B. Lyanda-Geller [a], M. J. Yang [b] and C. H. Yang [c]

[a] Department of Physics, Purdue University, West Lafayette, IN 47907

[b] Naval Research Laboratory, Washington, DC 20375

[c] Department of Electrical and Computer Engineering, University of Maryland

College Park, MD 20742

*Abstract*

We propose a novel scheme of solid state realization of a quantum computer based on single spin "enhancement mode" quantum dots (QD) as the building blocks. In the enhancement mode QD, just one electron can be brought into initially empty quantum dots, in contrast to the depletion mode QD based on expelling of electrons from multi-electron QDs by gates. The quantum computer architectures based on the depletion mode QDs are confronted by several challenges, making the scalability difficult. These challenges can be successfully overcome by the approach based on the enhancement mode capable of producing a lateral square array of quantum dots with versatile functionalities. These functionalities allow transportation of qubits, including teleportation and error correction based on straightforward one and two-qubit operations. We describe the physical properties and demonstrate experimental characteristics of the enhancement QD based on single-electron transistors (SETs) using InAs/GaSb composite quantum wells. We discuss the material aspects of the QD quantum computing, including advantages of materials with large spin splittings such as InAs and perspectives of enhancement mode approach in materials such as Si.

PACS: 73.21.La, 03.67-a, 81.07.Ta, 03.67.Lx

# I. Introduction

Quantum computers capable of solving important encryption and search problems with polynomial rather than exponential resources have attracted attention of researchers to quantum information processing.[1] Physical implementations of quantum computing have been actively searched for during the last decade. Numerous proposals for realization of qubits and quantum gates are available to date. One challenging issue is scaling quantum systems up to a large number of logical qubits, which consist of physical qubits and quantum gates. Within the time intervals less than the qubit dephasing time, i.e., while physical qubits maintain their coherence, they must be encoded into logical qubits via series of single-qubit and two-qubit operations. Physical qubits need to be initialized, manipulated, entangled, transported and read out. Two physical qubits coupled via exchange interaction realize the quantum gate. It has been proven that combining quantum gate operations, such as mutual spin flip of two qubits (i.e. swap) and single qubit spin rotations, is universal, i.e., sufficient for performing any quantum computation.[2]

Qubit transportation, a necessary procedure for encoding of logical qubits in ultimate quantum computer architectures, is an especially challenging aspect of solid state quantum computing. Transportation, single-qubit operation and entanglement of qubits require unprecedented degree of control of their quantum-mechanical wave functions.

Solid state and semiconductor implementations of qubits offer advantages of being scalable, evident from the modern semiconductor technologies. However, recent analysis [3] of certain solid state quantum computing schemes, e.g., based on Kane nuclear spin proposal,[4] have revealed fundamental problems in solid-state-based quantum computer design. There are limitations in control over qubits and in possibility to move them. It is therefore critical to develop flexible physical and architectural schemes that can overcome these challenges.

Among various scalable schemes that have been proposed as qubits, single electron in quantum dots (QDs) is a promising candidate.[5] Spin of a single electron in magnetic field provides a natural two-level system for realization of a qubit. Lateral quantum dots, in which spin detection is carried out by charge spectroscopy in the Coulomb blockade regime, are especially attractive because their properties are controlled by electrostatic gates.[6,7,8] A successful example of a gate-controlled system is GaAs/AlGaAs lateral QDs, in which single spin QDs and coupling of dots have been demonstrated.[9] Other solid state schemes that can benefit from advantages for scalability offered by semiconductor industry have not shown such coupling yet.

The key ingredient of the success of the lateral QD was the use of split electrostatic gates. These gates create lateral *depletion mode* single electron transistors (SETs), in which pairs of lateral tunneling barriers imposed by electrostatic gates define QDs.[9] The advantage of the lateral QDs is that coupling between qubits can be controlled by the same electrostatic metal gates defining QDs [10] which can be operated at sub-THz speed. This can allow millions of single-qubit and two-qubit gate operations performed within the qubit lifetime, i.e., spin dephasing time.[11, 12, 13, 14] However, lateral *depletion mode* SETs approach is unlikely to answer multiple challenges posed by requirements for quantum computing. First, it is difficult to get QDs with a single electron. In this system, the number of electrons in the QD is reduced from hundreds down to one by a plunger gate. Second, when the QD depletion is almost complete, there is an uncertainty in verifying that indeed only one electron remains in the dot and represents itself a qubit. The reason is that this identification is usually carried out by measuring the conductance through the dot. However, the plunger gate that reduces the numbers of electrons simultaneously raises the tunneling barriers and eventually shuts down the conductance.[15] Several electrons may remain in the dot in this case, preventing the dot from being a true single-spin qubit. To circumvent this drawback, an additional quantum point contact (QPC) adjacent to the SET was inserted to serve as an electrometer to verify that one electron remains in the dot.[8] The necessity for QPCs hinders the ability of scaling the system to a large number of qubits and quantum gates.

In contrast to the lateral QDs, another QD system, the vertical *depletion mode* InGaAs QDs, have well defined tunneling barriers determined by the conduction band offset with the AlGaAs layers.[16] As a result, there is no ambiguity in identifying one electron in the dot by measuring conductance in a vertical QD system. Unfortunately, in this setting, it is difficult to achieve sufficient coupling of vertical quantum dots.

We propose here a novel scheme for scalable QD quantum computing. In this scheme we use *enhancement mode* single electron transistors to create single spin quantum dots. We show the experimental data on enhancement mode QDs fabricated in composite InAs/GaSb quantum wells grown by molecular beam epitaxy, demonstrating the feasibility of the scheme. This approach possesses advantages of both lateral and vertical QD systems, including electrostatically controlled lateral configuration and well-defined tunneling barriers that allow direct identification of single spins in dots. Furthermore, it is straightforward to realize a square array of quantum dots underneath the electrostatic gates, with electric contacts to each of these dots for characterizing their conductance. QDs can be inserted in such an array or removed from there by gate voltages. Moreover, vertical electric gates allow the control of presence or absence of electrons on each dot, a physical transfer of electrons from an occupied dot to a neighboring empty dot, and coupling between two quantum dots each containing one single electron. As we shall see, this simple device layout makes it straightforward scaling up to a large number of logical qubits, transporting qubits and implementing error correction protocols. These protocols[1] consist of quantum operations on physical qubits and quantum gates that are able to identify errors and to perform quantum operations to correct the errors.

## II. Overview

We describe, in this paper, a concrete architecture design and detailed operations for initializing, manipulating, transporting, reading out information off physical spin qubits, and entangling them into physical quantum gates. We then discuss how to combine physical qubits and quantum gates, i.e., single spins in QDs and pairs of coupled dots, respectively, into logical qubits. Depending on the complexity of the error-correcting protocol, quantum information can be retained in a logical qubit with high probability well beyond the time scale set by the spin dephasing times. Logical qubits can then be combined into logical gates. In error-correction protocols, logical gates are encoded by a certain number of logical qubits, each of which can be encoded in turn by several physical qubits. This is called recursive encoding. Recursive encoding requires transporting of physical qubits or transporting the quantum information that they encode. As we shall describe, our scheme makes this possible by creating a gate-controllable 2D array of quantum dots with transporting spins within the array.

In the QD (SET) array, dots play versatile roles, and dots can be inserted or removed. Some of the dots can host electrons at a given time, while others can be used for measuring spins or for transfer of spins between different locations in the 2D array. In the majority of 2D semiconductor schemes, transfer of information is achieved via consequent swapping operations between neighboring qubits. This leads to an unavoidable difficulty of possible correlated errors: a qubit that is being transferred to some location on the array necessarily crosses paths of qubits that are swapped with another transferred qubit[1]. As we shall see, in our scheme simple transfer of spins via paths that contain empty quantum dots can significantly enhance the capability of transporting information within the system compared to swapping. This opens the possibility to prevent correlated errors.

Our paper is organized as follows. In Section **II,** we describe in detail the physics and the design of the enhancement mode SET and present the first experimental realization of such SET. We then discuss in section **III** the issue of materials for QD quantum computing and address the inherent advantages of materials with large g-factors such as InAs. In Section **IV,** we describe single physical qubit, initialization, manipulation, entanglement, and readout in system with enhancement mode SETs. In the same section we discuss quantum gates, composed of two coupled physical qubits. In section **V**, we describe transportation of qubits, including teleportation, which are necessary for fault-tolerant quantum computation. In Section **VI**, we discuss architectures using enhancement SETs for realization of error correction protocols. In section **VII**, we conclude and discuss generalizations of our approach, including opportunities arising with enhancement mode SETs in other materials, particularly in Si.

## III. Enhancement Mode Single Electron Transistor

The idea of the enhancement mode SETs, which can host one electron or be empty, with a single vertical electrical gate design is crucial in our scheme. Fig. 1 illustrates the contrast between "enhancement" and "depletion" mode for SETs and the conventional FETs (field effect transistors). The enhancement mode SET uses only one metal gate to define two tunneling barriers and create a single electron QD. This is in contrast to the lateral depletion mode SETs that have been pursued in the experiments on quantum computation in QDs so far. Depletion QDs require two sets of split gates to define a QD, and one additional Schottky gate to reduce the number of electrons in the QD down to one. Our scheme can be implemented using any narrow bandgap quantum well (QW) that can be biased from accumulation to inversion with a top metal gate. In this paper, we demonstrate the operating principle using an InAs/GaSb composite QW[17] (see Fig. 2 (a)). We will discuss the system based on InAs material properties and comment on the opportunities arising in materials such as Si in the conclusion section.

In the enhancement mode transistor, no electrons are present at zero applied voltage. When a positive voltage is applied to a single top metal gate, two symmetric tunneling barriers are created between GaSb and InAs QWs, as shown in Fig. 3 (c). These tunneling barriers define an InAs quantum dot and a single electron can tunnel there. As will be described in the following, identifying single electron spin in a QD poses no problem, in contrast to lateral depletion quantum dots. The important advantage of our vertical electrical gate design in the scheme is its similarity to the current practice in CMOS (complimentary metal oxide semiconductor) industry.

**III.1. The Design of Enhancement SET**

We now describe in detail the enhancement mode SETs that utilize the unique type-II broken gap band alignment between InAs and GaSb. Figure 2 (a) shows the schematic potential profile of the sample structure, which consists of an InAs/GaSb composite QW sandwiched by two barriers, e.g., $Al_xGa_{1-x}Sb$ as shown. The barrier is modulation p-doped so that there are two-dimensional (2D) holes in the GaSb first hole subband ($E_v$). The InAs QW is thinner than 7nm so that its first 2D subband energy ($E_c$) in equilibrium is significantly above the Fermi level of the system ($E_f$) and resides in the GaSb bandgap. In other words, the effective bandgap of this composite QW is tunable and can be less than 100meV. Sample preparation begins from the fabrication of 1D quantum wires containing charge-carrier holes with source and drain contacts, followed by oxide and metal evaporation to form the top gate (D-gate as the quantum Dot-gate), as illustrated in Fig. 2 (b). The overlap area between the hole wire and the D-gate defines the QD. The width of the conducting 1D hole channel (this width differs from the physical wire width by the two lateral depletion widths) and the D-gate width are around 100nm. Fig. 3 (a) shows the predicted trans-conductance characteristics and Fig. 3 (b) shows the trans-conductance characteristic that has been achieved experimentally to date. The schematic lateral potential profiles along the source-drain current direction are depicted in Fig. 3(c) for three different D-gate voltages. The source-drain current/conductance decreases monotonically as the D-gate voltage becomes more positive (regime I) in Fig. 3(b) and (c). It eventually vanishes when $E_v$ beneath the D-gate area is below the $E_f$ (regime II). Note that $E_c$ follows the same potential profile as $E_v$. The source-drain current remains zero when the $E_f$ lies between the $E_v$ and $E_c$ for the band profile beneath the D-gate area. However, when $E_c$ under the D-gate area is brought into resonance with the Fermi level, the InAs layer underneath the gate is ready to host electrons (regime III). Electrons populate this InAs QD through electron tunneling from GaSb layer. This is interband Zener tunneling. The barrier height for this tunneling is $\Delta E = E_c - E_v$. The tunneling amplitude is determined mainly by the InAs layer thickness.

Because of the smallness of InAs QD and the Coulomb blockade effect electrons tunnel only one at a time. As a result, the first conductance peak is directly associated with the lowest electron state in the InAs QD. In our scheme, we will use two kinds of dots: empty dots and dots occupied by a single electron.

**III.2. Materials issues**

We will now formulate optimal properties of semiconductor materials for quantum computing in QD systems. For this purpose, we discuss the example of InAs, the material that we used here for enhancement mode QDs.

InAs has a number of properties that are inherently advantageous for quantum information applications in comparison with other semiconductors, e.g., GaAs. Most importantly, InAs has a large g-factor ($|g^*|$=15 vs. 0.44 for GaAs)[18] so that Zeemann splitting of electron levels in InAs equal that in GaAs can be achieved with a magnetic field 30 times smaller. One of the important consequences of a larger g-factor is that it takes a smaller external microwave magnetic field to rotate spin via Rabi pulses (see Section **IV.3.**). In order to implement Rabi pulses, the period of Rabi oscillations in a physical qubit must be shorter than the spin dephasing time, $T_2$. Due to a larger $g^*$-factor, the constraint on the magnitude of ac magnetic field ($B_{ac} > \hbar(g^*\mu T_2)^{-1}$) is 30 times smaller for InAs compared to GaAs provided that $T_2$ are the same, which is feasible as we shall see. Consequently, the required ac electric current for generating $B_{ac}$ is 30 times smaller, which corresponds to power dissipation $10^3$ times smaller for qubit manipulation in InAs setting.

Another important property of the bulk InAs is the strong energy dependence of g-factor (compare 60 eV$^{-1}$ in InAs vs. 4.5 eV$^{-1}$ in GaAs).[19] This leads to Zeemann splittings in individual QDs different from dot to dot because of their different sizes that control the QDs energy spectra. Correspondingly, there is an opportunity for individual access to QDs via the respective a.c. magnetic field frequencies matching spin splitting of the QDs.

Another important property of InAs is that the orbital level energy spacing $\Delta E_{orb}$ in QDs is larger compared to GaAs due to a smaller electron mass (0.023m* vs. 0.067m*) and a steeper lateral confinement potential. Our experimental data[20,21] infer a 10meV sublevel spacing in 70nm wires. This large level spacing affects spin coherence times. The reason is that large energy spacing can reduce the level mixing through spin-orbit interactions. Therefore, despite stronger spin-orbit interaction in InAs compared to GaAs, the admixture of orbital states through this interaction can be relatively weak. Both the spin flip rate ($1/T_1$) and dephasing rate ($1/T_2$) are governed by such level mixing for principal spin relaxation and dephasing mechanisms, and a larger $\Delta E_{orb}$ leads to a relatively long spin coherence time. Moreover, large $\Delta E_{orb}$ brings the QDs closer to ideal two-level qubits.

Steep lateral potential profile in InAs quantum dots also means that two QDs can be placed in close proximity. This is desirable for enhancing the coupling strength in quantum gates in the regime when coupling is on (see Section **IV.4**).

We are now able to formulate the optimal properties for semiconductor materials for qubits:

- Large g-factor for smaller power dissipation
- Strong energy dependence of the g-factor in a bulk material to create different Zeeman splitting in dots of different size for individual qubit access.

- Large energy splitting between single-particle orbital levels for small admixture of spin states between different orbital levels in QDs (for long coherence times); and for making QDs closer to an ideal two-level system.
- Steep confinement potential for closer spacing (stronger coupling) between qubits

Finally, it is necessary that the material makes possible to construct an array of quantum dots each of which could be separately measured and controlled.

InAs-based structures that allow realization of enhancement mode SETs, provide the above-mentioned optimal features. Estimates of coherence times and dot couplings show that, using InAs QD as the building block, the spin logical gate can be realized. Another promising candidate for enhancement mode QDs is Si, which has a g-factor about 5 times larger than GaAs and a rather steep potential confinement. We now turn our attention to a description of the implementation scheme for quantum computer based on enhancement QDs.

## IV. Qubits Operations
### IV.1. Spin Qubit

With the state-of-the-art nanofabrication technique, it is currently possible to create InAs QDs with a diameter less than 100nm, limited by the physical width of D-gates. In perspective, we envision dots with sizes ~ 20nm. The smaller the dot size is, the larger $\Delta E_{orb}$ and the charging energy of the dot, $e^2/C$, will be. Here C is the total capacitance of the dot. Thus, the range of gate voltages in which InAs dots are occupied by single electrons is larger than in GaAs based system, opening stronger opportunities for modulating single-electron state by ac gate voltage.

One of the most important characteristics of the spin qubit is the spin dephasing time. Spin dephasing time determines whether the magnitude of $B_{ac}$ for spin manipulation is practical. It also sets the limits on the duration of gate operations. Spin qubit dephasing is determined by various mechanisms, such as hyperfine interactions of electron and nuclear spins and scattering by phonons in the presence of spin-orbit interactions. For conduction band electrons in III-V system such as InAs, it has been long appreciated that for dominant mechanisms, $T_1$ and $T_2$ are of the same order of magnitude. In a 2D electron gas, $T_2 = T_1/2$.[22,23,24,25] This property is the consequence of the symmetry of the intrinsic spin-orbit interactions in these systems. Recent analysis of relaxation and dephasing of lateral QD electrons by phonon scattering due to spin-orbit interactions has shown that this picture also describes localized electrons in QD.[26] However, there are experimental signatures[27] that effects of nuclei are important in QDs, and therefore relaxation and dephasing of qubits are to a large extent determined by hyperfine interactions of electron and nuclear spins. For fully isolated dots, hyperfine interactions can result in electron relaxation or dephasing, which is caused by electron-phonon interaction[12] or flip-flop processes of nuclear spins due to their dipolar coupling.[13] The important feature of the spin qubits is that in gated lateral dots, the fluctuations of gate voltage do not affect qubits directly, but rather through spin-orbit interactions. As a result, the estimates of the corresponding dephasing rates are at most in the range of the rates resulting from mechanisms, mentioned above. However, in Coulomb blockaded lateral dots, where electrons have final amplitude of tunneling off and on the dot, the resulting strength of interactions of electron and nuclear spins depends significantly on the applied gate voltage.[14] The gate voltage defines the escape rate of electrons from the dot, and therefore governs the effective magnetic fields, which result from interaction of spins of electrons in the dot with nuclear spins. These effective magnetic fields are also strongly influenced by electronic spin-orbit coupling.[14] Effective fields lead to electron spin relaxation and dephasing in Coulomb-blockaded QD, and therefore the electrostatic gate can effectively control qubit coherence. The viability of such control is an important feature, because operating a quantum computer requires changes in coupling of a given QD to the surrounding dots and to the regions of the 2D gas. Control of coherence can prevent errors caused by changes in coupling.

In the enhancement SET setting, one can control tunneling between the QD with a single spin and current leads by using a second top metal gate (I-gate as the isolation-gate), separated from the D-gate by a dielectric film, as illustrated in Fig. 4(a). Note that the confinement potential of the quantum dot is not affected by the electric field created by I-gate because of the screening of the D-gate metal. The ability of independent control of potential profile adjacent to the dot gives an additional leverage to maximize the spin lifetime.

As we have discussed, in general, $T_1$ and $T_2$ in InAs can be kept at the same level as in GaAs-based systems, due to the due to large energy spacing of orbital levels in InAs resulting in comparable admixture of spins due to spin-orbit interactions in these two materials. In particular, we estimate that the phonon-mediated electron spin dephasing[12,28] in 50nm InAs quantum dot at T=300mK is characterized by $T_2$=100 μs. Similar $T_2$ arises due to the effect of hyperfine coupling of electron and nuclear spins in the presence of dipole-dipole nuclear spin interactions.[13] Hyperfine interaction-induced decoherence of spins in the Coulomb blockade regime can result in shorter $T_2$, however, it can be tuned by gate voltage[14] to the same range of times as those defined by the phonon-mediated dephasing in the presence of spin-orbit interactions.

## IV.2. Spin Initialization

Initialization of qubits to the spin ground state is straightforward. In the presence of the external dc magnetic field parallel to the growth direction, a single spin-up electron enters the dot when the D-gate voltage brings the ground state dot level in resonance with the Fermi energy in the leads to the dot. It is noteworthy that the Coulomb blockade prevents another electron from entering the dot. Initialization of qubits to the ground state spin-up level is sufficient for starting the quantum computer, since a subsequent Rabi pulsing (as will be described in the next subsection IV.3) can bring qubits to any other state. If necessary, leads with the spin-polarized charge carriers characterizing the GaAs scheme[15] can be also used in our setup. The peculiar feature of the InAs/GaSb system is that charge carriers in the leads are the heavy holes. When these holes enter the dots they undergo interband tunneling transforming them into electrons. In magnetic field perpendicular to the 2D plane, holes can be spin-polarized at sufficiently low temperatures. The very useful feature of our setting is that the microwave in-plain magnetic field can couple only to electrons in the dot, but cannot couple to holes in the leads because their transverse hole g-factor is almost zero for [100] growth (the hole spin is pinned along the growth axis). Therefore, the spin polarization of holes in the leads is not be affected by the microwave magnetic field.

## IV.3. Manipulation of Qubits

Rabi pulsing, in our scheme, is intended for spin rotations and implementation of error-correction protocols. In order to make electron qubits capable of spanning the whole Bloch sphere, we need to use microwave excitation with arbitrary polarization, including the circular polarization. We propose to achieve such arbitrary polarization by using two current-carrying wires that produce microwave magnetic field parallel to quantum dot layer and are perpendicular to each other. As suggested by Vandersypen et al,[15] an integrated wire can be used to generate needed ac magnetic field with low power consumption.

In our scheme, D-gates and I-gates can be used themselves for generating two perpendicular microwave magnetic fields due to the vertical gate layout (as shown in Fig. 4 (b)). Thus, our transistor configuration naturally, without additional effort, adopts the scheme of Ref. [15] for creating ac magnetic fields. More specifically, we can superpose an ac voltage to the dc bias voltage applied to the D-gates (or I-gates) when y-direction (or x-direction) ac field is required. The number of electrons in the dot is stable within a wide range of D-gate and I-gate dc voltages under the condition depicted in Fig. 4(c), therefore it is feasible to modulate voltage on metal gates to generate local ac magnetic field. Furthermore, the additional advantage of our scheme is that though the D-gates can screen the dc electric fields generated by I-gates, it cannot screen out the ac magnetic field because the D-gate width (<100nm) is much smaller than the ac wavelength (>mm). Moreover, not only the amplitude of the required ac voltage is much smaller than that of the dc bias, but also it is possible to design a symmetric wire layout so that the voltage right at the dot location is not affected by the additional ac voltage. Due to a large $g^*$-factor in InAs, the required ac field is rather small, and the ac current running through D-gates is similarly small as well. If we assume $g^*=-10$ (we assume a smaller amplitude than the bulk value, due to the potential confinement), a period Rabi oscillations of 100ns, and a distance of 100nm between the D-gate metal and the InAs QD, the required ac field and the ac current are 71μT and 36μA, respectively. The corresponding ac voltage is 1.8mV (assuming a 50Ω terminator). That is small compared to the interval of dc voltage (i.e., spacing between the first and the second conductance peaks in Fig. 3a), which is of the order of 100mV, and therefore these parameters are realistic. As a result, the power consumption is remarkably low, just 46nW. This is almost three orders of magnitude lower than in analogous GaAs-based setup.

Due to the presence of two perpendicular ac magnetic fields in our scheme, spin rotation is achieved over any angle around arbitrary axis. Consequently, any one-qubit gate operation, in particular, the Hadamard gate, the Pauli X, Y and Z gates, as well as the phase gate and the $\pi/8$ gate can be executed on a single qubit. By adding an additional exchange two-qubit gate via coupling of the QDs, as will be described in the following subsection, we make any two-qubit gate operations possible.

### IV.4. Coupling of Two Qubits

In quantum computing settings it is necessary to have a mechanism for switching on and off coupling ("on" and "off") between two qubits. When coupling is off, individual qubits are manipulated, and qubit can be effectively isolated from the environment preventing decoherence and loss of quantum information. Some residual coupling can be present in the "off" duration, and it is important that its value is minimized. When coupling is on, quantum information is transferred between the qubits. However, in order to prevent dephasing of both spins in coupled qubits, it is necessary to keep the "on" duration sufficiently short. .

The enhancement SET configuration offers a straightforward solution for achieving controllable coupling between two neighboring qubit spins by inserting an additional metal gates (J-gates in Fig. 4(a)). Two physical mechanisms of coupling can be realized depending on the configuration of the two dots at a given instance of time. These mechanisms supplement each other. One mechanism is the direct tunnel coupling between the two spins, which is carried out by lowering the barrier between the two dots with the J-gate. The schematic potential profile along the x-direction is plotted in Fig. 5. Note that because the orbital level spacing in QDs is sufficiently large, sub-THz pulses will have almost no effect on individual qubit spins, and will be important for exchange interaction only. The distance between centers of dots that can be achieved by state-of-the-art fabrication is 30-40 nm. (Currently, the distance between dots that can be realized in laboratories is 100 nm). In these regimes, tunneling amplitude between dots is much smaller than the Coulomb repulsion U on the dot. Therefore, lowering the barrier does not lead to double occupation of dots that could constitute an error in computation. In the coupling "on" regime, the magnitude of the exchange interaction is about J=5 μeV, assuming the distance between centers of the dots is 100nm, $\Delta E_{orb}$=10 meV, and the barrier height between the two dots is 5meV. For the barrier height 20 meV, the coupling is J= $5 \cdot 10^{-9}$ eV (coupling "off" regime).

The other method of creating controllable coupling of spins on two dots is using the J-gate to create an intermediary dot with an electron that would mediate indirect exchange coupling of qubit electrons. This approach substantially suppresses the tunneling amplitude between two qubits $t_G$, thereby providing means to further reduce residual coupling of qubits in the coupling "off" regime. In this case, the corresponding exchange energy is of the order of $t_G^2/U$. For coupling between qubit dot and intermediary dot, we have $t_i^2/U$, where $t_i$ is the corresponding tunneling amplitude. The estimate for the indirect coupling of qubits is $t_i^4/(U^2 \Delta E_{in})$, where $\Delta E_{in}$ is the energy difference between ground states of qubits and intermediary dot. The indirect exchange is comparable to direct exchange for U=2meV, $\Delta E_{orb}$=10 meV, and $\Delta E_{in}$ =0.1 meV in the coupling "on" regime.

A mixed approach can be implemented by creating an intermediate dot without an electron; with additional voltage pulse, electron spin from qubit dot can tunnel into this dot, interact, e.g., swap their spins, and return back to it original qubit dot. Combining two-qubit swap operations with single-qubit rotations, physical gates performing other logic operations such as C- NOT and XOR can be realized.

**IV.5. Read-out**

Qubit readout in our scheme is achieved by converting electron spin to charge due to spin-dependent tunneling on and off the dot in the Coulomb blockade regime. That means that one SET and one QD are responsible for the readout. In brief, readout SET, tuned to a Coulomb blockade peak, senses if an electron is present on the readout dot R. Figure 6 sketches a small scale quantum computer comprising of the enhancement dots, where each row represents a logic qubit and each column represents a logic gate. Such 2D array of spin qubits resembles a single-electron-charge-couple device.

In magnetic field, tunneling from the qubit dot to the dot R is spin-selective. With strong dependence of InAs QD g-factor on the size of the dot, making dot R with a larger g-factor compared to g-factor of the qubit dot will allow tunneling from the qubit dot if the qubit spin is in the ground state, which is close in energy to the dot R ground spin state. However, tunneling of the qubit electron with the opposite spin will not be allowed, because electron energies for this spin state on two dots differ significantly. The readout starts with a voltage pulse that transfers an electron from the qubit dot to the dot R. The charge of dot R changes if qubit electron is in the ground state. This change of the charge is then measured. Similarly to the GaAs system, the estimates show that at this stage it is feasible to transfer electron from the qubit dot to the dot R within 100ps and to measure it by SET within 1ns.

## V. Moving Qubits

### V.1. Swapping channel

Any large-scale quantum computing requires communication of quantum information between different parts of the device. In contrast to classical information, quantum information cannot be copied (because of the quantum non-cloning theorem[1]) and must be transported from its source to destination. Possibilities to transport quantum information that have been discussed to date are to move quantum data using swap operations (swapping channel) or to use teleportation.

In solid state setting for quantum computer, swapping is envisioned between nearest neighbor qubits. The sequence of nearest neighbors forms the swapping channel, in which the quantum information is progressively swapped between pairs of qubits. Challenges for swapping in the quantum computer based on phosphorous atoms in Si have been analyzed in[3]. The principal challenges are (i) difficulties of placing qubits in proper locations, (ii) different dimensions of qubits and control lines complicating fabrication, (iii) possible quantum behavior of control lines, which need to be classical, and (iv) appearance of correlated errors. Quantum dot arrays offer significant advantages over phosphorous atoms or other nuclear spin qubits schemes. First, in a QD array scheme there is no hurdle of placing QDs, which are sufficiently large and are defined by metallic D-gates. Second, control lines into quantum data-paths have the same spatial dimensions as QDs, and therefore fabrication problems have been largely solved. Third, control lines are metallic, and no problem of quantum behavior arises on the 20nm spatial scale there. The reason is that metallic structures with sizes comparable to sizes of QD exhibit no quantum mechanical effects because the electron Fermi wavelength in metal (sub-nm) is much smaller than the physical dimension. In addition, no problem of "connecting" classical wires to atomic size qubits[3] arises if qubits are as large as QDs.

We now estimate the latency time and the bandwidth (the amount of information that is possible to transfer during 1s and the time it takes to traverse the line of qubits, correspondently), for a swapping channel without error correction in the QD array scheme. Having in mind that only every other QD in our QD array will be used as a qubit, we take the separation between qubits 100nm. With 5µeV exchange coupling in the "on" regime, the duration of the swap operation is $\sim 10^{-10}$ s. Therefore for 1µm wire, the latency time is $t_{lat}$=1ns, three orders of magnitude smaller than for phosphorous atom in Si scheme. Furthermore, the physical bandwidth is 1/1ns=$10^9$ quantum bits per second. It is three orders of magnitude more than in the scheme using qubits based on P nuclei in Si. The true bandwidth is proportional to fidelity that is defined as

$$f=\exp(-\lambda d),$$

where d is the length of the wire measured in number of qubits, and $\lambda$ is the ratio of the characteristic spin dephasing time to the duration of the swap operation, $\lambda=10^{-6}$ in our scheme. Therefore, the true bandwidth for 1µm wire is 99999x$10^4$ bits per second. However, similarly to the Si scheme discussed in Ref. [3], this calculation of the bandwidth is correct as long as the fidelity remains above the threshold $10^{-4}$ required for fault-tolerant quantum computation. This gives us the maximum distance of the swapping channel

$$d=d_0 \ln(1-10^{-4})/(-\lambda),$$

which is 1μm, or 10 qubits for our system. Thus, the prospects of using swapping operations for transportation of qubits are quite limited.

**V.2. Tunneling channel**

Our 2D array of enhancement scheme offers an alternative mechanism for transporting qubits and quantum information. Note that QDs in the array can carry different functionalities and, in particular, can contain one electron or none of them. We envision that by filling only every other QD with electron we can create a path, in which qubit electrons will move from filled to empty QDs. The sequence of QDs that consists of one filled and several empty QDs form the tunneling channel.

The importance of tunneling channel is two-fold. As we shall see in the next sections, tunneling channel enhances capabilities of our scheme for teleportation and error correction. Furthermore, because qubits transported along tunneling channel are not swapped with other qubits, an error in a moving qubit does not increase errors in other qubits, and correlated errors are avoided. This is a significant improvement over other solid state architectures that use only swapping channels for moving qubits.

The time that takes to move charge from occupied dot to an empty dot is defined by the amplitude of tunneling between dots. The characteristic tunneling energy scale is an order of magnitude bigger than the exchange interaction, and time necessary to move charge is at least an order of magnitude shorter than time needed for swapping operation. Therefore, the maximum distance for tunneling channel exceeds 10μm, or $10^2$ qubits. However, large-scale quantum computation requires larger number of qubits. To achieve an architecture that involves $10^5$-$10^6$ qubits, teleportation techniques discussed in Ref. [3] are needed.

### V.3. Teleportation channel

A system of enhancement QDs offers opportunities of effective teleportation of qubits. Quantum teleportation is the recreation of a given quantum state at a distance. Using classical communication lines that transmit measured values of qubits without full quantum information of the states provides relaxed requirements for operating the system. In brief, the teleportation procedure is the following (Fig. 6): (i) on a preliminary stage, an entangled Einstein-Podolsky-Rosen (EPR) pair of qubits is created, and one qubit |A> is kept at the source while the other qubit |B> is fully transmitted to a destination via swapping or tunneling channels; (ii) qubit |C> that is to be sent via the teleportation channel interacts with the source EPR qubit |A> via CNOT operation. Subsequently, the phase of qubit |C> and the amplitude of qubit |A> are measured and their value are sent via classical lines to a destination; (iii) next, the measured amplitude of qubit |A> controls the application of the Pauli gate X to qubit |B> and then, the result of measurement of phase of qubit |C> controls the Pauli gate Z applied to qubit |B>. These operations recreate the state of qubit |C> by transforming the state |B> into it, while the original state of qubit |C> has been destroyed by measurements, in full accord with the non-cloning theorem.

The EPR pair needed for teleportation is created in our scheme by using two neighboring QDs. First, qubits are initialized (sec. IV.2) in their ground states |0>. Then, e.g., the Hadamard gate is applied with the help of electrical J- and D-gates to the first qubit, bringing it to a superposition $(|0> + |1>)/\sqrt{2}$. Finally, the CNOT two-qubit operation is performed on the second qubit and the superposition $(|0> + |1>)/\sqrt{2}$, resulting in entangled two-qubit state $(|00>+|11>)/\sqrt{2}$.

On the preliminary stage of the teleportation, quantum transmission of the EPR pair qubit |B> is still a necessary step. However, in this case a certain (known) qubit is transmitted, which is much more advantageous compared to the transmission of the unknown quantum information contained in |C>. Because qubits |A> and |B> have known properties, a purification procedure[29,30,31] can be applied in order to allow transmission of EPR qubit |B> to a distance exceeding tunneling channel length. The purification is achieved by applying only local gates to qubits |A> and |B>, and using classical communication lines for coordinating these local gate operations.

Purification results in a fraction of pure EPR pairs that are ideal for use in teleporting. Due to the possibility to use tunneling channels, both purification and the teleportation itself are significantly more effective in our scheme than for technological schemes used as example in Ref. [3]. The bandwidth of the teleportation process takes into account the effectiveness of purification. We estimate that the true bandwidth of the 1cm-long line of qubits in InAs coupled QD array scheme is $\sim 1.65 \times 10^8$ bits per second. For teleportation, the size of the array is essentially not limited. For longer teleportation distances, purification must discard more and more corrupted EPR pairs. This effect, however, does not pose any problems for quantum computing.

## VI. Logical Qubits and Error-Correction

Similar to quantum computation itself, logical qubit and implementing error-correction protocols use combined single-qubit and two-qubit operations. Creation of logical qubits is rather straightforward in our scheme, with current capabilities of technology and fabrication that make possible in-plane arrays of coupled spin qubits combined with readily available readout. In particular, simple error correction schemes combine just a few qubits. The smallest number of qubits that can readily implement a simple error-correction protocol is five.[32] Steane logical qubit capable of error correction [33] includes seven qubits, and Shor's simple scheme stable against arbitrary errors uses nine qubits.[34] Five-qubit and Steane logical qubits in our coupled array can be represented by one-dimensional chains of quantum dots, which can be coupled via direction transverse to chains for forming logical gates.

Simple five-qubit error correction scheme consists of one principal qubit, which value is to be protected, and four syndrome qubits in initial state spin down. These five qubits are encoded in one logical qubit by a set of unitary operations. Then, logical qubit can couple to other logical qubits for performing the quantum computation. Each physical qubit comprising the logical qubit and particularly the principal qubits are subject to decoherence, which produces errors. At any moment, the state of the logical qubit can be checked by decoding the state with unitary transformation. Decoding unitary transformation is the inverse to encoding unitary transformation. The properties of the unitary transformation guarantee that measured states of the syndrome qubits determine which error occurred. It is important in this regard that measuring syndrome qubits does not destroy the quantum information on the principal qubit that is not coupled to syndrome qubits in the decoded state. When the type of the occurred error is detected, one is able to determine the operation (which is a combination of single-qubit Pauli gates) to be performed on the principle qubit. The encoding operation can than be repeated to resume the quantum computation.

Gating operations by electrostatic metal gates in our scheme can proceed with sub-terahertz speed. That means that in the five-qubit error correction scheme, qubit encoding, decoding and error correction sequences of pulses that includes in total about 500 pulses, can be performed $\sim 10^4$ times during the spin dephasing time.

All principal elements of error correction, such as creating and un-creating "cat states" of the form $(|0000\ldots>+|1111\ldots>)/\sqrt{2}$ that allow avoiding transmission of errors through CNOT gates, and parity measurements, give the possibility for our scheme to develop error-correction codes more complicated than 5-qubit error correction protocols. Using the capability to move qubits via tunneling and teleportation channels, the implementation of multiple logical gates, computing on encoded qubits, concatenation of qubits and recursive error correction are only a matter of improvement of technology and fabrication, and have no fundamental limitations.

## VII. Conclusion

We have presented a scheme for scalable quantum computation in QD array. The main feature of this scheme is very low power dissipation, and the ability to move and teleport QD qubits via tunneling channels that prevent correlated errors. The key element of the scheme is the enhancement mode SET approach that allows straightforward creation of multifunctional elements, including creation and removal of empty dots or electrons in them. We have demonstrated enhancement dot SET in InAs/GaSb system. The direction of the work in the near future is to confirm that relatively small size of QDs gives desired spin relaxation and dephasing times of the order of tenth of a millisecond as estimated by theory and to compare them with the existing data in GaAs system, where SET technology have been demonstrated in recent years.

Similar design principle based on the enhancement approach can be applied in silicon based metal-oxide-semiconductor structures. Using a patterned top gate, single electrons can be induced or depleted at the MOS interface. Single spins created in silicon or silicon-germanium alloy have several desirable properties useful for quantum information processing. In silicon, single electrons are expected to have very long spin coherence time because the hyperfine interaction with the nuclear spin is very weak (because the dominant naturally occurring isotope Si 28 has zero spin). [35] Electrons bound to donors in Si have a spin relaxation time $T_1$ of hours. Single electrons at ground state confined in quantum dots are expected theoretically to have equally long coherence time, because the spin-orbit interaction is very small in MOSFETs and can be further suppressed in Si-Ge QDs. The g-factor is smaller in Si and Si-Ge systems compared with InAs, but it is ~5 times bigger than in GaAs system. This opens the possibility to reach dissipation levels 20 times smaller than in GaAs. Advantages of long coherence times and unprecedented level of development of Si technology makes Si systems promising candidates for scalable quantum computing. Experiments aimed at creating enhancement mode SET in Si QDs are currently under way.

## VII. Acknowledgements

We are grateful to I. L. Chuang for useful discussions. Supports from NSA/ARDA, LPS/NSA and ONR are gratefully acknowledged. YLG acknowledges supports from NRL under contract No N00173051G013.

**Figure Captions**

Figure1: Schematics of (a) the depletion mode field effect transistor; (b) the enhancement mode field effect transistor. The left figures show the structure of the transistors, and the right show the expected transfer current-voltage characteristics.

Figure 2: (a) Schematic band diagram of the composite quantum well. (b) Schematic of the enhancement-mode single electron transistor, where one electron is induced by the "D-gate."

Figure 3: (a) Expected transconductance of the enhancement-mode single electron transistor. (b) Measured drain current versus a sweeping gate voltage at a constant drain voltage of 1mV, at 4.2K. (c) The potential profile along the source-drain direction, showing the InAs conductance band minimum (top curve) and the top of the GaSb valence band (bottom curve). The dashed line depicts the Fermi level.

Figure 4: (a) Schematic showing the structure of the single electron transistor with the "D-gate" and the "I-gate." (b) Side view of the same transistor structure, showing the directions of the two perpendicular magnetic fields created by pulsing currents through the D- and the I-gates. (c) Schematic showing the placement of a "J-gate" for controlling the coupling of qubits.

Figure 5: (Left) Potential profile of two isolated qubits, with no voltage applied to the J-gate. (Center) Coupled qubits with a positive voltage applied to the J-gate. (Right) Indirect coupling of qubits via an intermediate qubit as discussed in the text.

Figure 6: Two dimensional array of qubits.

Figure 7: (a) Filled and empty dots in a quantum dots array and tunneling path; (b) Scheme of teleportation of qubit |c⟩, where. |a⟩ and |b⟩ form an EPR pair.

Figure 1

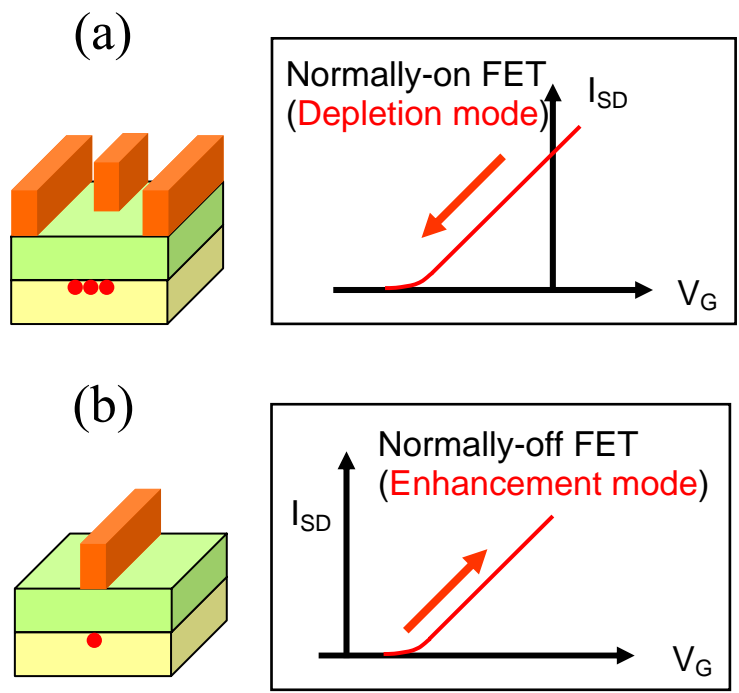

Figure 2

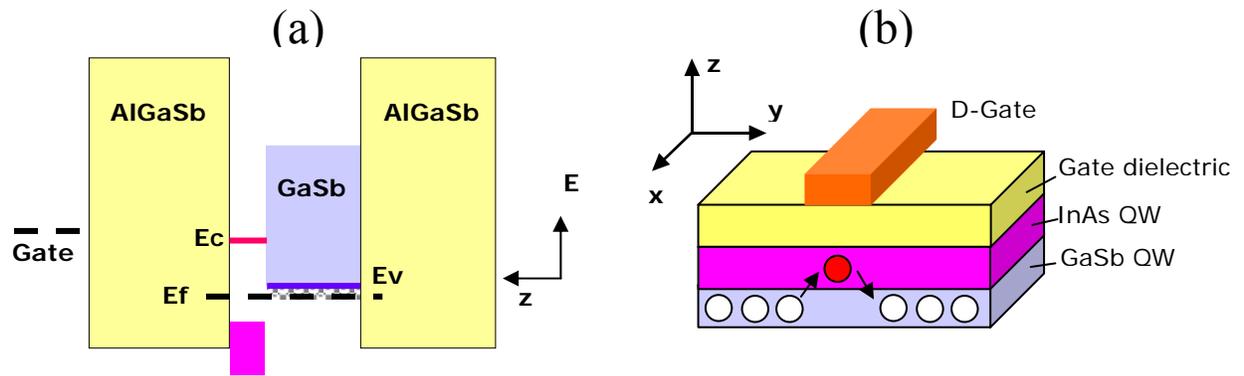

Figure 3

(a) 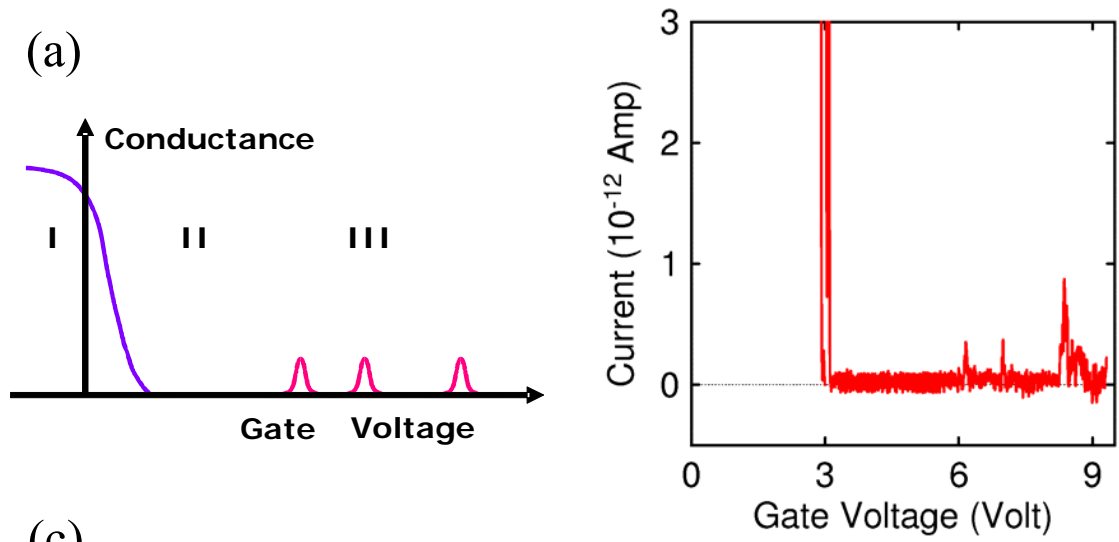

(c)

I. Accumulation 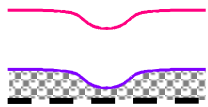

II. Depletion 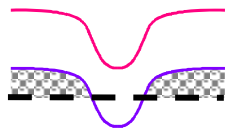

III. Inversion 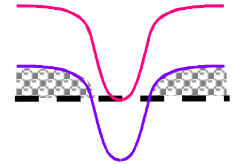

Figure 4

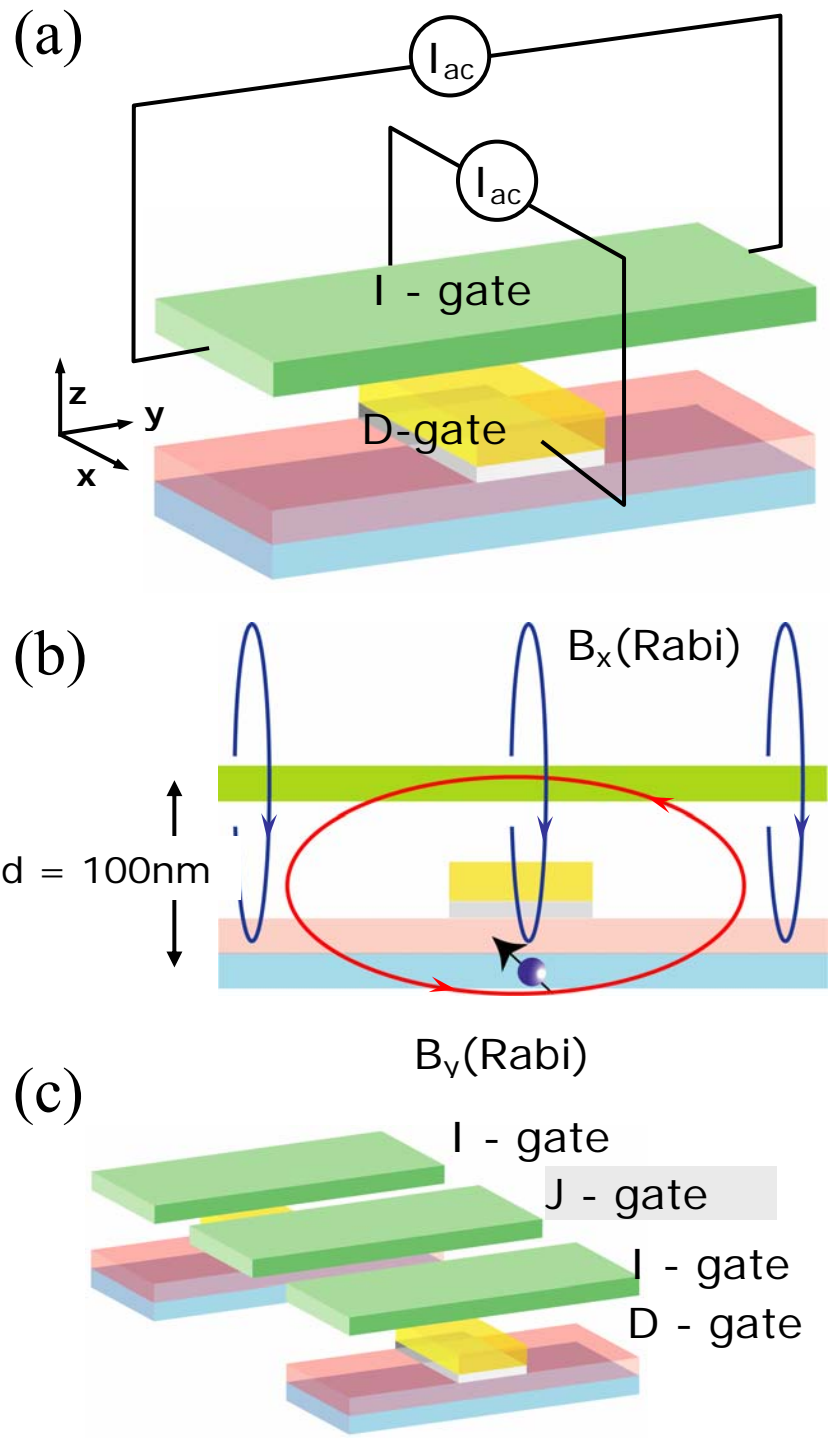

Figure 5

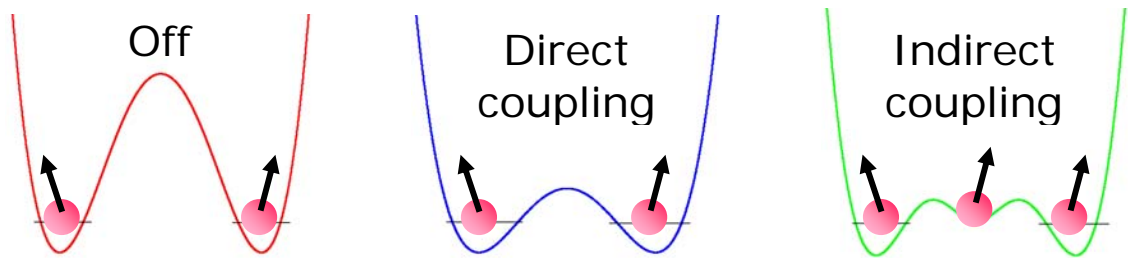

Figure 6

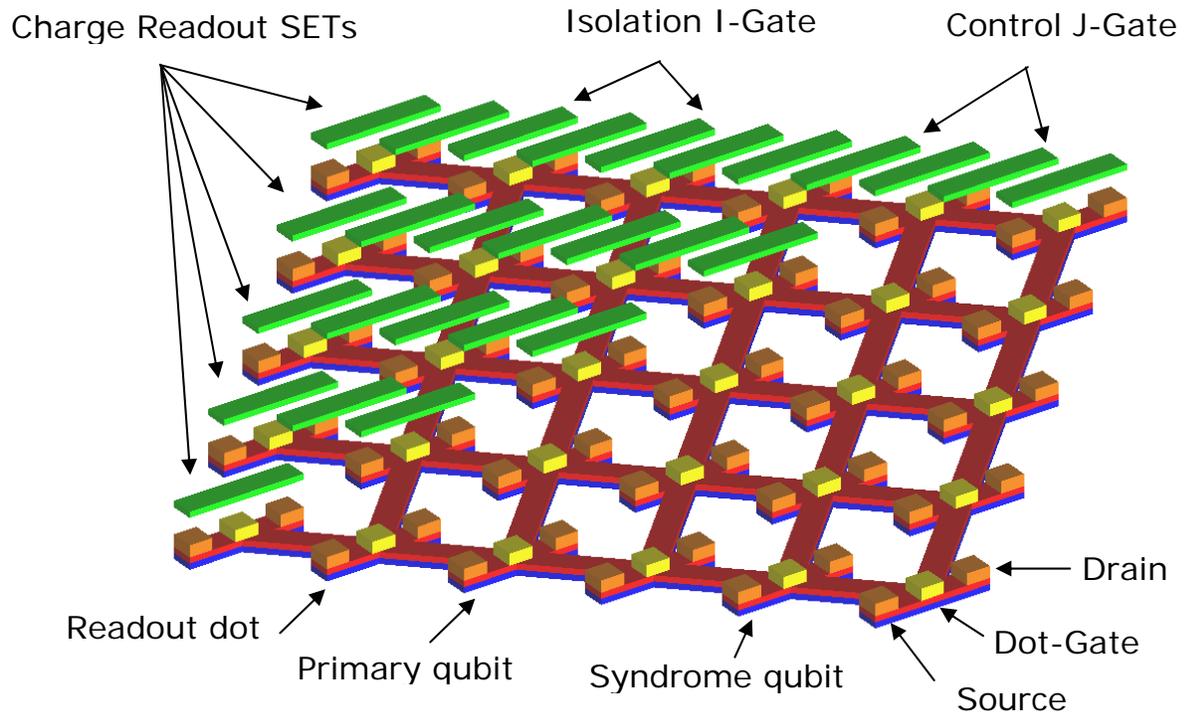

Figure 7

(a)

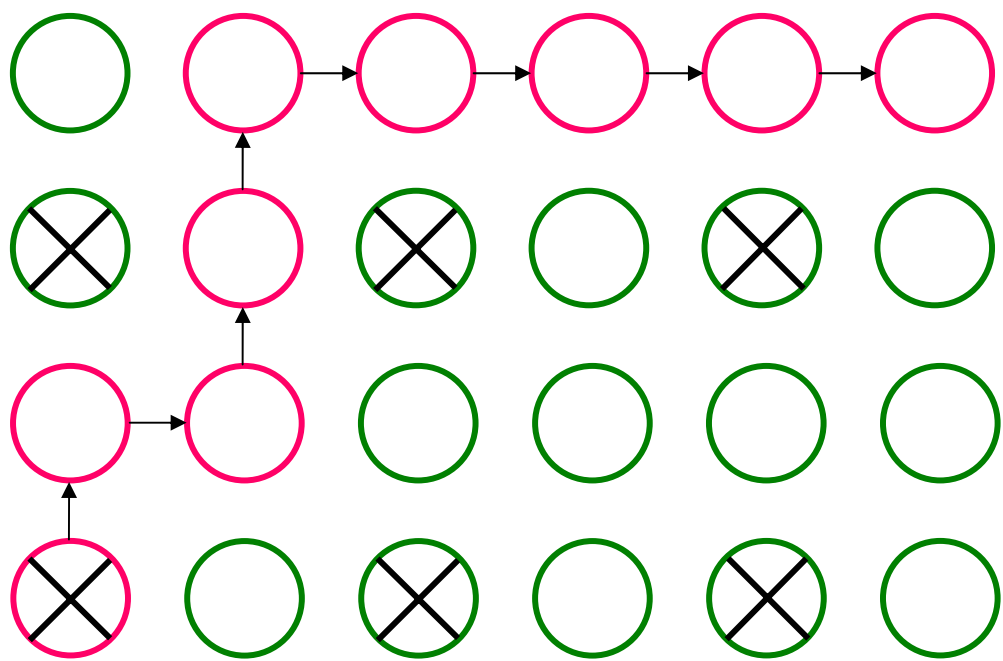

(b)

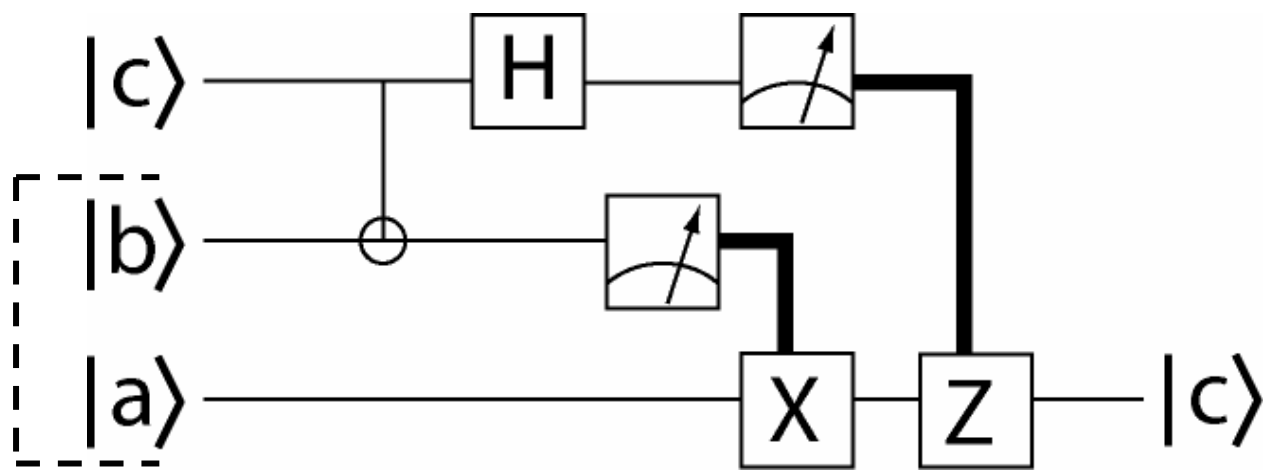

**References:**

[1] M. A.Nielsen and I. L. Chuang, Quantum Computation and Information, Cambridge University Press Cambridge UK, (2000).

[2] D. DiVincenzo, Phys. Rev A 51, 1015 (1995).

[3] D. Copcey, M. Oskin, F. Impens, T. Tetodiev, A. Cross, F. T. Chong, I.L. Chuang and John Kubiatowicz, IEEE J. Selected Topics in Quant. Electronics, 9 1552 (2003).

[4] B. E. Kane, Nature 393 133 (1998).

[5] D. Loss and D. P. DiVincenzo, Phys. Rev. A, 57 120 (1998).

[6] L. P. Kouwenhoven, C. M. Marcus, P.L. McEuen, S. Tarucha, R. M. Westervelt and N.S. Wingreen, in Mesoscopic Electron Transport, Vol 345 of NATO Advanced Study Institutes, Series E: Applied Sciences edited by L.L. Sohn, L.P. Kouwenhoven and G. Schoen (Kluwer Academic, Dordrecht, 1997) p. 105.

[7] M. Ciorga, A. S.Sachrajda, P. Hawrylak, C. Gould, P. Zawadzki, S. Jullian, Y. Feng and Z. Wasilewski, Phys. Rev. B 61, R16315 (2000).

[8] D. Sprinzak, Y. Li, M. Heiblum, D. Mahalu, and H. Shtrikman, Phys. Rev. Lett. 88, 176805 (2002).

[9] J. M. Elzerman, R. Hanson, J. C. Greidanus, L. H. Willems van Beveren, S. De Franceschi, L. M.K. Vandersypen, S. Tarucha and L. P. Kouwenhoven, Phys. Rev. B 67, 161308 (2003).

[10] G. Burkard, D. Loss and D. P. DiVincenzo. Phys. Rev. B 59, 2070 (1999).

[11] A.V. Khaetskii and Y.V. Nazarov, Phys. Rev. B 64, 125316 (2001); LM Woods, T.L. Reinecke and Y. B. Lyanda-Geller Phys. Rev. B 66, 161318 (2002).

[12] Erlingson, V. I. Falko and Y. V. Nazarov, Phys. Rev. B 64 195306 (2001).